\documentclass[pre,aps,reprint,amsmath,amssymb,longbibliography,superscriptaddress,groupedaddress,floatfix]{revtex4-2}
\usepackage{amsmath}

\usepackage[pdftex]{graphicx}       
\usepackage{txfonts}        
\usepackage{mathrsfs}	
\usepackage{orcidlink}	
\usepackage{layout}

\newcommand{\revision}{}

\newcommand{\Ffrac}{\left(\frac{-F}{k_\text{B} T}\right)}
\newcommand{\eref}[1]{Eq.~\eqref{#1}}
\newcommand{\fref}[1]{Fig.~\ref{#1}}
\newcommand{\sref}[1]{Section \ref{#1}}

\DeclareMathOperator{\var}{var }
\usepackage{hyperref} 
\hypersetup{colorlinks,citecolor=blue,filecolor=blue,linkcolor=blue,urlcolor=blue}

\begin{document}

\title{Compressed self-avoiding walks in two and three dimensions}

\author{C J Bradly\orcidlink{0000-0002-5413-777X}}
\email{chris.bradly@unimelb.edu.au}
\author{N R Beaton\orcidlink{0000-0001-8220-3917}}
\affiliation{
	School of Mathematics and Statistics, University of Melbourne, Victoria 3010, Australia
}
\author{A L Owczarek\orcidlink{0000-0001-8919-3424}}
\noaffiliation


\begin{abstract}
We consider the phase transition induced by compressing a self-avoiding walk in a slab where the walk is attached to both walls of the slab in two and three dimensions, and the resulting phase once the polymer is compressed. 
The process of moving between a stretched situation where the walls pull apart to a compressed scenario is a phase transition with some similarities to that induced by pulling and pushing the end of the polymer. 
However, there are key differences in that the compressed state is expected to behave like a lower dimensional system, which is not the case when the force pushes only on the endpoint of the polymer. 
We use scaling arguments to predict the exponents both of those associated with the phase transition and those in the compressed state and find good agreement with Monte Carlo simulations.
\end{abstract}

\maketitle


\section{Introduction}
\label{sec:Intro}

The effect of confinement on polymers is of great interest in statistical mechanics.
The development of atomic force microscopy and optical and magnetic tweezer methods have enabled the manipulation and confinement of individual polymers \cite{Neuman2008}.
A related question is the effect of confinement on biopolymers, either for translocation of biopolymers or pharmaceutical compounds through pores and nanotubes \cite{Slonkina2003,Sotta2000} or how a polymer escapes out from under the compression of an AFM tip \cite{Guffond1997}.
\revision{
In complex fluids polymers can be adsorbed to colloidal particles in scenarios like steric stabilisation where there is an effective interaction between colloids \cite{Verma1998} or with a wall \cite{Milchev2002}. 
When the colloidal particles are large and close together, the adsorbed polymers can be modeled as confined to an infinite slab and subject to compressive forces.
Similarly, a single polymer chain in a network can be viewed as confined and the properties of single subchains subject to stresses may be important for the bulk properties of polymer gels \cite{Shirai2023}.
}

In statistical mechanics the self-avoiding walk (SAW) is the canonical model for understanding the configurational properties of long polymer molecules \cite{Madras1993,Rensburg2015} and possesses the same scaling properties as real polymers \cite{DeGennes1979,Edwards1965}.
We consider SAWs as paths on a $d$-dimensional hypercubic lattice $\mathbb{Z}_d$ that do not visit the same site twice.
If the number of such SAWs of length $n$ is denoted $c_n$ then it is known that
the growth constant $\log \mu = \lim_{n \to \infty} n^{-1} \log c_n$ exists \cite{Hammersley1957}, and thus it is generally believed that asymptotically $c_n \sim A \, \mu^n n^{\gamma - 1}$ for some universal exponent $\gamma$.
In order to consider walks compressed between two parallel surfaces, we are interested in the subset of SAWs that are restricted to lie in the upper half-space, with one end attached to an impermeable surface.
If $c_n^+$ counts such walks then similar relations hold as for $c_n$, with the same growth constant $\mu$ \cite{Hammersley1982} but $\gamma_1$ replaces $\gamma$ \cite{Guttmann1984}.
\revision{
	For the stronger restriction where the walks are confined between two walls that are at a fixed distance apart, lattice polymers have long been considered a model of steric stabilisation \cite{Guttmann1978,Brak2005,Rensburg2005}. 
	Rigorous results for the limiting free energy of self-avoiding walks between two walls indicate the presence of a repulsive entropic force by the polymer acting on the confining walls \cite{Rensburg2006}.
}
However, it has proven difficult to obtain rigorous results about SAWs that go far beyond these asymptotic relations, which motivates the use of numerical simulation for comparison to theoretical arguments.

It is worth mentioning that there are many other lattice models that are exactly solvable and for which the effect of an applied force can be incorporated.
These models are based on directed or partially-directed walks; the directedness \revision{often} makes the models solvable but often makes them less realistic.
However, the latter is less of an issue when the system is already anisotropic due to geometric confinement or application of a force.
For a review of these models see \cite{Orlandini2016}.
\revision{In two dimensions  the model of directed walks that are confined to a slab but without an applied force can be solved exactly \cite{deBruijn1972}.
In the case of a compressive force this is more difficult but the generating function and the partition function scaling has been found exactly \cite{Guttmann2015}.
It is not clear whether a suitable directed model exists in three dimensions since solvable directed models in three dimensions are often pseudo two-dimensional \cite{Brak2010}. 
In any case, our focus here is to ascertain the correct scaling exponents for the SAW model.}

For self-avoiding walk models the force is typically considered to be applied in one of two ways, resulting in similar situations \cite{Rensburg2009a}.
First, the force can be applied to the endpoint and the average height of the endpoint is the appropriate order parameter.
When $F > 0$ the walk is pulled taut and is in the ballistic phase, meaning the endpoint height scales linearly with $n$ \cite{Ioffe2010}.
It is also known rigorously that the point of zero force, $F = 0$, is a critical point between the ballistic and extended phases \cite{Beaton2015}. 
\revision{At this critical point the critical exponents $\phi$ (related to the finite-size crossover) and $\alpha$ (related to the specific heat scaling) have expected values} $\phi = \nu_d$ and $\alpha = 2 - 1/\nu_d$ \cite{Bradly2023}, where $\nu_d$ is the Flory exponent for $d$-dimensional SAWs.
In particular, since $1/2 \leq \nu_d < 1$ for $d \geq 2$, then $\alpha < 1$ and hence the ballistic-extended transition is continuous.
In the pushed phase of this model $F < 0$, a small pushing force towards the impermeable surface only has a local effect on the endpoint of the walk, and in the limit $F \to -\infty$ the SAW effectively becomes a loop with both ends confined to the surface, but the walk is otherwise free to explore the bulk.

Alternatively, the model we consider here is where the force is applied to \revision{the walls to expand or compress the confining space, with the condition that the walk is always in contact with both surfaces.}
Then the average span of the walk $s$, or size of the gap between the surfaces, replaces the endpoint height as the order parameter of interest.
This model has been considered previously \cite{Beaton2015b}, but only for two dimensional walks, relying on the conjectured existence of a conformally invariant scaling limit for SAWs \cite{Lawler2004}.
The point of zero force $F = 0$ is still expected to be a critical point, but the properties of the transition have not been thoroughly explored.
\revision{Because the walk is neither tethered to the upper surface nor interacting with it, the $F > 0$ phase is less physically relevant and we concentrate on the compressive $F < 0$ side where the effects of confinement are potentially more interesting.}
This phase has not been studied as extensively, especially for $d = 3$, \revision{where the scaling derived from 2D SAWs (or any directed model) does not directly apply.}
When $F < 0$ in the compressed phase it is intuitive that the confining surface will induce quasi-($d - 1$)-dimensional behaviour \cite{Vanderzande1998,Micheletti2011}, but we will see that this is not the case for both $d = 2$ and $d = 3$ dimensions.
Furthermore, the arguments of Ref.~\cite{Bradly2023} indicate that when the force is applied at the endpoint, the span of the walk scales as $n^{\nu_d}$; this is different to the scaling of the span when the force is applied to another surface located at span $s$.

In \sref{sec:Model} we define the model of compressed walks and Monte Carlo simulations used to simulate the problem.
In \sref{sec:Compressed} and \sref{sec:Critical} we discuss the scaling behaviour in the compressed phase and critical point, respectively.
Finally in \sref{sec:Results} we compare these scaling arguments to numerical estimates of exponents for both 2D and 3D cases.

\section{Model of compressed walks}
\label{sec:Model}

\begin{figure}[t!]
	\centering
	\includegraphics[width=\columnwidth]{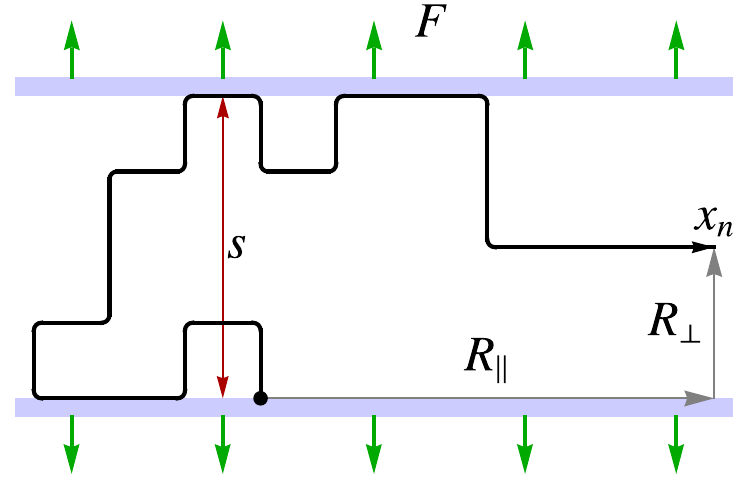}
	\caption{
	A self-avoiding walk on the square lattice confined to a strip of width $s$ and a \revision{positive (negative) force $F$ is applied to the walls to expand (compress) the walk.} 
	The walk is terminally attached to the lower surface and touches the upper surface at least once but the endpoint $x_n$ is otherwise free within the strip.
	\revision{The position of $x_n$ gives the components of the endpoint distance $R$.}
		}
	\label{fig:ExampleCompressedSAW}
\end{figure}

Our model is of self-avoiding walks on $\mathbb{Z}_d$, especially $d = 2,3$, confined between two parallel impermeable surfaces.
\revision{One end of the walk is tethered to the bottom surface and the highest parts of the walk, not necessarily the other end, is assumed to always be in contact with the upper surface.
A force is applied to expand or compress the walk between the surfaces,} see \fref{fig:ExampleCompressedSAW}.
The height of the highest vertices and thus the height of the upper surface is the span $s$ of the walk.
The partition function for such walks is
\begin{equation}
	Z_n(F) = \sum_s a_{ns} y^s,
	\label{eq:PartitionFunction}
\end{equation}
where $a_{ns}$ is the number of walks of length $n$ and span $s$, and $y = \exp\left( F/k_B T \right)$ is the Boltzmann weight for a force $F$ applied to the top surface.

For $y > 1$ ($F > 0$) the walk is stretched between the two surfaces and for $y < 1$ ($F < 0$) it is compressed.
The order parameter indicating these two phases is the average (fractional) span
\begin{equation}
	s_n(F) = \frac{\langle s \rangle}{n} = \frac{1}{n \, Z_n(F)} \sum_s s \, a_{ns} y^s,
	\label{eq:AverageSpan}
\end{equation}
and its variance (specific heat) is
\begin{equation}
	c_n(F) = \frac{\var(s)}{n} = \frac{\langle s^2 \rangle - \langle s \rangle^2}{n}. 
\label{eq:SpecificHeat}
\end{equation}
Although the force is not applied to the endpoint, we are still interested in the behaviour of the endpoint, namely the end-to-end distance $R$.
Because of the geometry of the model it is useful to further distinguish between the components of $R$ parallel and perpendicular to the surface where the walk is attached, namely 
\begin{equation}
	R_\parallel = \sqrt{\left\langle x_{n,1}^2 + \ldots + x_{n,d-1}^2 \right\rangle}
	\quad \text{and} \quad
	R_\perp = \sqrt{\left\langle x_{n,d}^2 \right\rangle},
\label{eq:RDefinitions}
\end{equation}
respectively, and $x_{n,i}$ is the $i^\text{th}$ component of the endpoint\revision{, see \fref{fig:ExampleCompressedSAW}.}

Generally, we are interested in the finite-size scaling of the metric quantities because the value of the scaling exponents is indicated by the phase, including at the critical point.
Thus we suppose, to leading order, that the metric quantities have large $n$ asymptotic scaling
\begin{equation}
	s_n \sim A_s n^{\nu_s - 1}, \quad R_\parallel \sim A_\parallel n^{\nu_\parallel} \quad \text{and} \quad R_\perp \sim A_\perp n^{\nu_\perp},
	\label{eq:MetricScaling}
\end{equation}
for non-universal constants $A_s, A_\parallel, A_\perp$ that are independent of $n$.
At $F = 0$ we expect $\nu_s = \nu_\perp = \nu_\parallel = \nu_d$, the Flory exponent for $d$ dimensions.
In the stretched and compressed phase these exponents may deviate from each other.

\subsection{Monte Carlo simulations}
\label{sec:Simulations}


We simulate SAWs in a strip or slab with the flatPERM algorithm \cite{Prellberg2004} that approximately enumerates the number of self-avoiding walks $a_{ns}$.
This algorithm samples walks by starting at the origin and growing the endpoint using a pruning and enrichment strategy and measuring the span $s$.
The main output is a flat histogram $W_{ns}$ that approximates $a_{ns}$, for all lengths $n$ up to a preset maximum length $n = 1024$. 
Note that during the simulation the walks are not confined to a strip and the effect of the force applied to the span of the walks is applied by weighting with $y$, according to \eref{eq:PartitionFunction}.
For each of $d = 2, 3$ we initiated 10 independent instances of flatPERM, each instance uses a parallel implementation with 8 threads, and each thread runs $10^5$ iterations.
This obtains a total of $\approx 1.2 \times 10^{10}$ samples at maximum length $n = 1024$.
The parallel implementation speeds up the initial equilibration phase \cite{Campbell2020} and incurs insignificant overhead since the parameter space is large enough for threads to avoid collisions.
In addition to the histogram $W_{ns}$ for the number of samples, the simulations also output histograms that accumulate the weighted endpoint position $W_{ns} x_n^2$ for calculating $R_\parallel$ and $R_\perp$.

\begin{figure}[t!]
	\centering
	\includegraphics[width=\columnwidth]{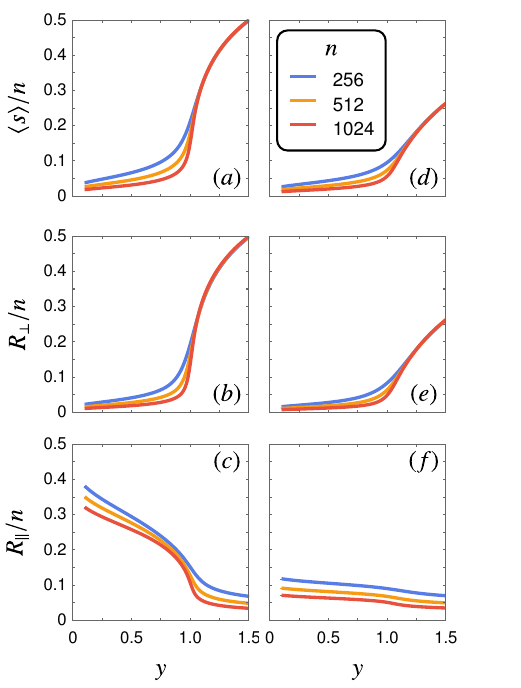}
	\caption{
	Metric quantities for compressed SAWs for (\revision{left}) $d = 2$ and (\revision{right}) $d = 3$.
	From \revision{top to bottom} are the average extension above the surface $\langle s \rangle$, the average height of the free endpoint, and the average in-plane distance of the free endpoint, each given as a fraction of length $n$.
	Curves are shown for several different lengths $n = 256, 512, 1024$. 
	}
	\label{fig:Metrics}
\end{figure}

From the simulation outputs we show in \fref{fig:Metrics} the metric quantities $s_n$, $R_\perp / n$ and $R_\parallel / n$ \revision{(top to bottom)} as a function of Boltzmann weight $y$. 
The endpoint components are scaled by $n$ for comparison to the order parameter $s_n$.
The \revision{left} plots are for walks on the square lattice ($d = 2$) and the \revision{right} plots are for walks on the simple cubic lattice ($d = 3$).
In the stretched phase, $y > 1$ ($F > 0$), the average span and distance of endpoint from bottom surface, $R_\perp$ are positive; they will asymptote to unity as $y$ increases.
In the compressed phase, $y < 1$ ($F < 0$), these quantities are approximately 0 in the large $n$ limit.
For both phases these quantities are larger in two dimensions than three dimensions, but otherwise are qualitatively the same.
Conversely, the component of the endpoint parallel to the confining surfaces $R_\parallel$ is suppressed in the stretched phase for both two and three dimensions.
However, in the compressed phase $R_\parallel$ remains small for $d = 3$ but is positive for $d = 2$.
This suggests that a two-dimensional walk obtains an effective directedness as it is increasingly confined to a narrow quasi--one-dimensional strip.

\section{Scaling in the compressed phase}
\label{sec:Compressed}

When the force is directed downwards ($F < 0$) the walks are compressed between the two surfaces.
Intuitively, one might expect that as $y \to 0$ (increasing compressive force $F \to -\infty$), the walk becomes effectively $(d-1)$--dimensional.
In which case, the in-plane endpoint size $R_\parallel$ scales like a free self-avoiding walk on a $(d-1)$--dimensional lattice.
The out of plane quantities are suppressed but behave as a random walk, meaning $\nu_s = \nu_\perp = 1/2$ for $F \ll 0$, but the amplitudes of $s_n$ and $R_\perp$ tend to zero.
However, when $F < 0$ but $|F|$ is not large the walks are in a compressed phase but still explore a $d$-dimensional space and other scaling occurs.

It is strongly believed, although not strictly proven, that two-dimensional self-avoiding walks have a conformally invariant scaling limit in which case the measure on such walks must be the $\mathrm{SLE}_{8/3}$ measure \cite{Lawler2004}.
In particular, for walks restricted to the upper half-plane and tethered to the surface, the number of walks, and thus the measure, scales as $n^{\gamma_1}$, where $\gamma_1$ is the entropic exponent.
It is predicted that $\gamma_1 = 61/64$ \cite{Guttmann1984} for two-dimensional walks tethered to a surface and this matches the properties of the corresponding SLE measure.

In two dimensions, the assumption of the scaling limit allows the following argument for SAWs restricted to a thin strip, following \cite{Beaton2015b}.
Consider SAWs restricted to a narrow strip of height $s$, such that the length is $n = K s^{1/\nu}$ for some large $K$.
These walks are a concatenation of $K$ subwalks, each of length $s^{1/\nu}$ where each smaller subwalk spans the strip.
Assuming the scaling limit exists, the lattice spacing can be rescaled to length $1/s$ in which case the SAWs can be viewed as quasi-1D walks with step size 1 and length $K$.

Then the measure of walks within a narrow strip of width $s$ can be viewed as the product of the measure of 1D walks and the measure of half-space walks
\begin{equation}
	q_n^{(\text{strip }s)} = q_K^\text{(1D)} q_n^+.
	\label{eq:MeasureStrip}
\end{equation}
The measure of walks with span \emph{exactly} $s$ is the derivative of this measure
\begin{equation}
	q_n^{(\text{span} = s)} = \frac{\partial}{\partial s} q_n^{(\text{strip } s)}.
	\label{eq:MeasureSpan}
\end{equation}
To relate this measure to our statistical mechanical model, note that the measure on a set of walks is related to the number of such walks by the growth constant $\mu$, that is $a_n = q_n \mu^n$.
If we also weight each walk by a Boltzmann factor \revision{$y^s$ for a compressive force $F < 0$ ($y = \exp(F / k_B T) < 1$)} and integrate over the span $s$ then we have the leading order term (for large $n$) of the canonical partition function \eref{eq:PartitionFunction}.
Hence for SAWs when $F < 0$ we have
\begin{align}
	Z_n(F) &\sim \mu^n \sum_{s} e^{F s / k_B T} q_n^{(\text{span} = s)} \nonumber \\
	&\sim \mu^n \sum_{s} e^{F s / k_B T} \frac{\partial}{\partial s} q_K^\text{(1D)} q_n^+.
	\label{eq:PartitionMeasure}
\end{align}
The measure of 1D walks is simply $q_K^\text{(1D)} \sim A_1 \exp\left( - \beta_1 K \right)$ and the measure of half-space walks is $q_n^+ \sim A_2 n^{\gamma_1 - 1}$, for some constants $A_1$, $A_2$ and $\beta_1$.
Also, recall that $n$ and $K$ are related by $n = K s^{1/\nu}$ and $K$ is a large constant in the scaling limit.
Using these expressions in \eref{eq:PartitionMeasure}, and a calculation of the asymptotic behaviour of the resulting integral (see Appendix A of \cite{Beaton2015b}), we arrive at the partition function for SAWs in the compressed phase
\begin{align}
	Z_n(F) &\sim A \, \mu^n \Ffrac^{\frac{\frac{3}{2} - \gamma_1 }{\nu + 1}} n^{\frac{\gamma_1 - 1 + \nu/2}{\nu + 1}} \exp\left( - \lambda \, n^{\frac{\nu }{\nu + 1}} \Ffrac^{\frac{1}{\nu + 1}} \right),
	\label{eq:PartitionCompressedGeneral}
\end{align}
where $A$ and $\lambda$ are constants independent of $n$ and $F$.
Compare \eref{eq:PartitionCompressedGeneral} to the expected scaling of free half-space self-avoiding walks \revision{$Z_n^+ \sim A^+ \mu^n n^{\gamma_1 - 1}$}; for compressed walks there is a larger leading order power, but an additional exponential suppression, indicating slow convergence.

From \eref{eq:PartitionCompressedGeneral} we can calculate the average span in the compressed phase \cite{Rensburg2016}
\begin{align}
	s_n(F) &= \frac{k_\text{B} T}{n} \frac{\partial \log Z_n}{\partial F} \nonumber \\
		&\sim \frac{\lambda}{\nu + 1}  \Ffrac^{-\frac{\nu}{\nu + 1}} n^{-\frac{1}{\nu + 1}} - \frac{\gamma_1 - 3/2}{\nu + 1} \Ffrac^{-1} n^{-1}.
	\label{eq:AverageSpanCompressedGeneral}
\end{align}

For $d = 2$ we have known values of the exponents $\nu = 3/4$ and $\gamma_1 = 61/64$ \cite{Guttmann1984}, thus the partition function is explicitly
\begin{equation}
	Z_n^\text{(2D)}(F) \sim A \, \mu^n \left(\frac{-F}{k_B T}\right)^{5/16} n^{3/16} \exp \left( -\lambda \, \left(\frac{-F}{k_B T}\right)^{4/7} n^{3/7} \right),
	\label{eq:PartitionCompressed2D}
\end{equation}
and the (fractional) average span is
\begin{align}
	s_n^\text{(2D)}(F) 
		&\sim \frac{4 \lambda}{7 n^{4/7}} \Ffrac^{-3/7} - \frac{5}{16n} \Ffrac^{-1},
	\label{eq:AverageSpanCompressed2D}
\end{align}
which indicates that $\nu_s = 3/7$ is the leading order exponent in the compressed phase. 

For the end-to-end distance scaling we note that the perpendicular component of the endpoint is bounded by the span of a walk, thus $R_\perp / n < s_n$.
However, the scaling is the same and we expect $\nu_\perp = \nu_s$.
For the parallel component $R_\parallel$, we consider again that a compressed 2D SAW is a quasi-1D SAW, meaning the compression induces a directedness.
and the height of the endpoint can be considered as a 1D simple random walk on top of the directed walk.
Hence we would expect $R_\perp \sim \sqrt{R_\parallel}$.
Since we already have $\nu_\perp = \nu_s$ this implies $\nu_\parallel = 2\nu_s = 2\nu_d/(\nu_d + 1)$ in the compressed phase.

\begin{table}[b!]
\centering
	\begin{tabular}{cc|c|c|c|c}
			& 	& $F \to -\infty$	& $F < 0$ & $F = 0$ & $F > 0$ \\ 
		\hline
		general & $\nu_s = \nu_\perp$	& -- 		& $\frac{\nu_d}{\nu_d + 1}$ & $\nu_d$ & $1$ \\ 
				& $\nu_\parallel$		& $\nu_{d-1}$ 	& $\frac{2\nu_d}{\nu_d + 1}$ 		& $\nu_d$ & $1/2$ \\
		\hline
		$d = 2$ & $\nu_s = \nu_\perp$	& -- & $3/7$ & $3/4$ & $1$ 	\\ 
				& $\nu_\parallel$		& $1$ 	& $6/7$ & $3/4$ & $1/2$	\\
		\hline
		$d = 3$ & $\nu_s = \nu_\perp$	& -- & $0.370\ldots$ & $0.588\ldots$ & $1$ 	\\ 
				& $\nu_\parallel$		& $3/4$ & $0.740\ldots$ & $0.588\ldots$ & $1/2$	\\
	\end{tabular}
	\caption{Conjectured metric exponents in each phase. It is important to note that the $d=3$ values are speculative and $\nu_\parallel$ does not match the numerical results. The limit $F\to-\infty$ indicates that the slab has zero width and the exponents $\nu_s$ and $\nu_\perp$ are not defined.} 
\label{tab:Exponents}
\end{table}

While for $d = 2$ the SLE scaling limit is not proven but it is strongly believed to be correct, for $d = 3$ there is no equivalent conjectured scaling limit.
However, even if we could rescale the walk in 3D, the quasi-2D walks in the compressed phase are not self-avoiding, and remain as 3D SAWs, except when the slab has zero width.
Nevertheless, it is intuitive that for a thin slab the quasi-2D walks are weakly self-avoiding and so the above argument may be approximately correct even for 3D.
For higher dimensions certainly we expect that $\nu_s$ is less than the 2D value in the compressed phase since there is more room for the walk to spread out when compressed.
The 3D exponents $\nu$ and $\gamma_1$ are known to high precision, so we can still test \eref{eq:PartitionCompressedGeneral} for $d = 3$.
We can say even less about the end-to-end distance scaling. 
We still expect $\nu_\perp = \nu_s$ for $d = 3$, but since the compression does not induce a directedness to the walks we do not have a prediction for $\nu_\parallel$ in the compressed phase.

Table \ref{tab:Exponents} summarises the expected values of the metric exponents in all phases, both generally and for $d = 2$ and $d = 3$. Explicit values for two and three dimensions are based on known values of $\nu_2 = 3/4$ and $\nu_3 = 0.587597(7)$ \cite{Clisby2010}.
The limit $F \to -\infty$ indicates a $(d-1)$--dimensional system where the slit or slab has zero width.
The $F > 0$ regime is the ballistic phase where the span and endpoint height scale linearly with $n$ and the parallel component is a small simple random walk.
The transition at zero force is discussed more in the next section.
For the compressed phase $F < 0$ it is important to note that the conjectured values for $d=3$ are speculative.
In \sref{sec:Results} we do find numerical agreement with the $d=3$ conjecture for $\nu_s = \nu_\perp \approx 0.370$, but we \emph{do not} find numerical agreement with the $d=3$ conjecture for $\nu_\parallel \approx 0.740$. 


\section{Scaling at the critical point}
\label{sec:Critical}

This model has a critical point at $F = 0$, between stretched and compressed phases.
The scaling at this critical point is expected to be similar to the endpoint-pulled model, examined recently \cite{Bradly2023}.
Like that model, we have standard crossover scaling \emph{Ans\"atze} for a continuous transition near a critical point $F_\text{c}$ at large length $n$, namely
\begin{equation}
	s_n(F) \sim n^{\alpha \phi - \phi} \tilde{S}\left( \left[ F - F_\text{c} \right] n^\phi \right), \mbox{ as } n \to \infty \mbox{ as } F \to F_\text{c},
	\label{eq:SpanCrossover}
\end{equation}
and
\begin{equation}
	c_n(F) \sim n^{\alpha \phi} \tilde{C}\left( \left[ F - F_\text{c} \right] n^\phi \right), \mbox{ as } n\to \infty \mbox{ as } F \to F_\text{c},
	\label{eq:SpecificHeatCrossover}
\end{equation}
where $\tilde{S}$ and $\tilde{C}$ are scaling functions and $\phi < 1$ is the crossover exponent.
The exponent $\alpha$ is defined by scaling in the thermodynamic limit where the specific heat 
$C(F) = \lim_{n \to \infty} c_n(F)$ scales as
\begin{equation}
	C(F) \sim \left( F - F_\text{c} \right)^{-\alpha}  \mbox{ as } F \to F_\text{c}.
\label{eq:AlphaDefinition}
\end{equation}
In order to reconcile the thermodynamic and large $n$ limits, standard tricritical scaling \cite{Brak1993} applies, whereby there is a relation between exponents $\phi$ and $\alpha$
\begin{equation}
	2 - \alpha = \frac{1}{\phi}.
	\label{eq:HyperscalingRelation}
\end{equation}
Thus at the critical point $F_\text{c} = 0$ we have
\begin{equation}
	s_n(0) \sim S_0 \, n^{\phi - 1} 
	\label{eq:SpanScalingAtF0}
\end{equation}
and 
\begin{equation}
	c_n(0) \sim C_0 \, n^{2\phi - 1}.
	\label{eq:SpecificHeatScalingAtF0}
\end{equation}
The constants $S_0$ and $C_0$ are the values of the scaling functions $\tilde{S}$ and $\tilde{C}$ at $F = 0$.

\begin{figure}[t!]
	\centering
	\includegraphics[width=\columnwidth]{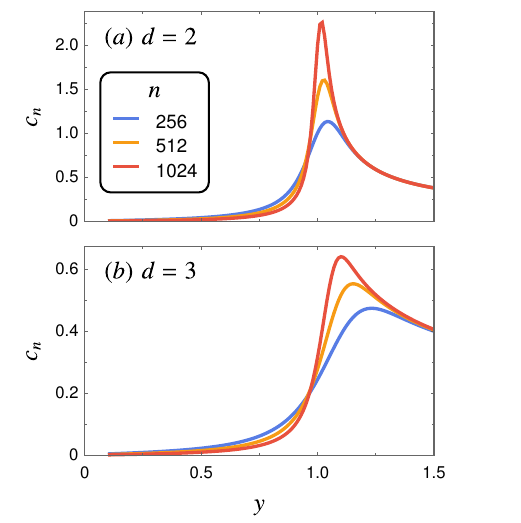}
	\caption{
	Specific heat $c_n$ for compressed SAWs, for (a) $d = 2$ and (b) $d = 3$.
	Curves are shown for several different lengths $n = 256, 512, 1024$, indicating that a peak is forming at the critical point $y = 1$ ($F = 0$).
	}
	\label{fig:SpecificHeat}
\end{figure}

Similar to the model with force applied to the endpoint, the span of a walk is a metric quantity, as discussed with \eref{eq:MetricScaling}. 
In particular, when there is no applied force, the span exponent should be the Flory exponent $\nu_d$ of a free self-avoiding walk confined to a half-space.
Thus at the critical point $F = 0$ comparison to \eref{eq:SpanScalingAtF0} indicates that $\phi = \nu_d$.
By the hyperscaling relation \eref{eq:HyperscalingRelation}, and the fact that $\nu_d \geq 1/2$ we also find that $\alpha < 1$, indicating a continuous transition.
Furthermore, the quantity $2\phi - 1$ is therefore positive (for $d = 2, 3$) and so the variance of $s$ at the critical point, namely $c_n(0)$, increases with $n$.
As seen in \fref{fig:SpecificHeat}, this manifests as a peak in the finite-size specific heat which serves as a good signature of the critical point.
If we assume that the peak is located very near to $F = 0$ as $n$ increases then the height of the peak scales as $n^{\alpha \phi}$ which is a reliable way to numerically estimate the exponents $\alpha$ and $\phi$.
The other way to estimate $\phi$ is to directly fit the average span at zero force, and obtain $\phi=\nu_s$ from \eref{eq:SpanScalingAtF0}.
An estimate for $\alpha$ is then derived from the hyperscaling relation \eref{eq:HyperscalingRelation}.

\section{Numerical estimates of exponents}
\label{sec:Results}

\subsection{Testing asymptotic expression for average span}
\label{sec:TestingExactForm}

\begin{figure}[t!]
	\centering
	\includegraphics[width=\columnwidth]{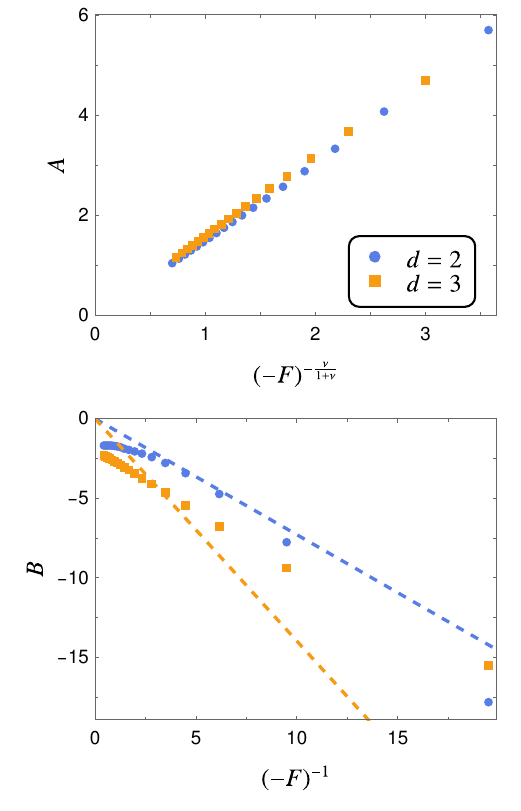}
	\caption{Estimates of coefficients of the exact expression for $s_n$ in the compressed phase ($F < 0$), \eref{eq:AverageSpanCompressedGeneral}. The coefficient $A$ of the leading order term behaves as expected, showing simple linear dependence on the known power of $F$, but the coefficient $B$ of the next order term does not. The dashed lines have slopes $(\gamma_1 - 3/2) / (\nu_d + 1)$.
	}
	\label{fig:ExactCoefficients}
\end{figure}

The scaling arguments for the compressed phase resulted in an expression for the average span, \eref{eq:AverageSpanCompressedGeneral} that not only indicates scaling as length $n$ increases, but also that the coefficients have specific expressions that depend on the compressive force.
There is a free parameter $\lambda$ but otherwise we can test the simple model $s_n \approx A / n^\frac{1}{\nu_d + 1} - B / n$ and compare to \eref{eq:AverageSpanCompressedGeneral}\revision{, assuming force $F$ is in thermal units, or $k_B T = 1$}.
In particular, we would expect $A \propto (-F)^{-\frac{\nu_d}{\nu_d + 1}}$ and $B \propto (-F)^{-1}$.
In \fref{fig:ExactCoefficients} we plot estimated parameters $A$ and $B$ against the appropriate inverse powers of $(-F)$, for values of $F < 0$ in the compressed regime.
Clearly, $A$ is proportional to $(-F)^{-\frac{\nu}{\nu+1}}$ as suggested by the scaling arguments.
Furthermore, a simple estimate of the slope gives estimates $\lambda \approx 2.626(1)$ for $d = 2$ and $\lambda \approx 2.536(2)$ for $d = 3$.
The value for $d = 2$ matches the estimate from \cite{Beaton2015b}, which was obtained from analysis of the exact enumeration of two-dimensional walks confined to a slit.
However, while the leading order term is supported by the numerical evidence, the data for the second term coefficient $B$ is not.
The dashed lines indicate the expected slope $(\gamma_1 - 3/2)/(\nu_d + 1)$, but the data is not even linear with respect to $(-F)^{-1}$.
A more careful analysis indicates that this fails because the next-to-leading order scaling of $s_n$ is not $1/n$, which indicates that the higher order terms are not so simply captured by the exact expression given by \eref{eq:AverageSpanCompressedGeneral}.

\subsection{Finite-size scaling analysis}
\label{sec:FSSResults}

\begin{figure}[t!]
	\centering
	\includegraphics[width=\columnwidth]{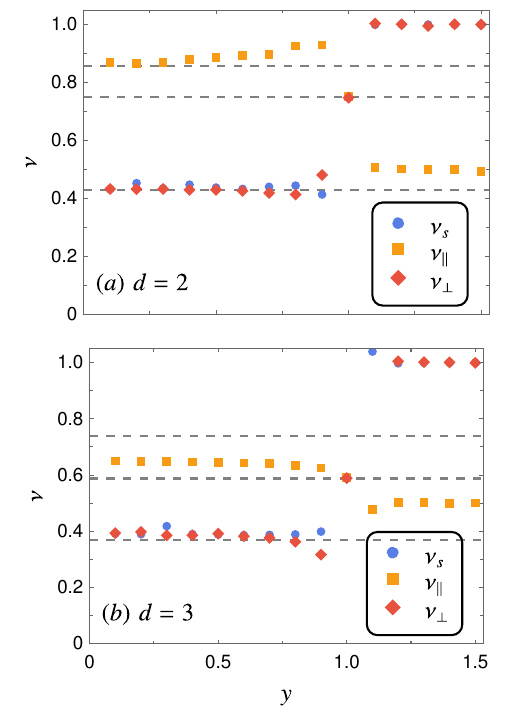}
	\caption{Exponent estimates for metric exponents $\nu$ as a function of pulling weight $y$, for (a) $d = 2$ and (b) $d = 3$.
	From bottom to top, the dashed lines mark the values of $\nu_d/(\nu_d + 1), \nu_d, 2\nu_d/(\nu_d + 1)$, where $\nu_d$ is the known exponent for free SAWs, namely $3/4, 0.587\ldots$ for $d = 2, 3$, respectively.
	}
	\label{fig:Exponents}
\end{figure}

If the predicted expression for $s_n$ is only partially correct, then we can fall back on standard methods for finite-size scaling properties.
For fixed $y$, we next estimate the leading order scaling exponents $\nu_s, \nu_\perp, \nu_\parallel$, as defined by \eref{eq:MetricScaling}.
In practice, we extend these scaling relations to include a correction-to-scaling term, which accounts for higher-order terms that may still be relevant at the lengths accessible by simulation, but without fixing the second order term explicitly as in the previous analysis.
Thus, for example the data for the average span is fitted to 
\begin{equation}
	s_n(F) \sim S \, n^{\nu_s - 1} \left( 1 + \frac{\tilde{S}}{n^\Delta} \right),
	\label{eq:SpanCorrectionToScaling}
\end{equation}
where the coefficients $S$ and $\tilde{S}$ are independent of $n$ but may vary with $F$.
As we have seen above the higher order terms are not well known, but it is necessary that an additional correction to scaling term is included.
The correction to scaling exponent is chosen to be $\Delta = 1/2$ to capture higher order terms, which is based on known estimates of higher order terms in the expansion of metric quantities of SAWs.
For example, the end-to-end distance of free SAWs in three dimensions has next order term with exponent $\Delta = 0.528$ \cite{Clisby2010}.
In the compressed phase $\Delta = 1/2$ is close to the predicted exponent for the second order term discussed above, but here we are agnostic about the precise value and the coefficient of the higher order terms, instead focusing on an accurate estimate of the leading order exponent.
Furthermore, this analysis can be applied for any $y$, not just the compressed phase, and can also be used to estimate $\nu_\perp$ and $\nu_\parallel$ from the end-to-end distance $R$.
Using correction-to-scaling forms, we estimate the metric exponents, shown in \fref{fig:Exponents} over a range of $y$ for (a) $d = 2$ and (b) $d = 3$.
In both two and three dimensions, the critical point at zero force ($y = 1$) is clear whereby all metric exponents take the value $\nu_d$ of free self-avoiding walks (marked by the middle gray dashed line).

In the stretched phase ($y > 1$) there is clear ballistic behaviour where the span and the height of the endpoint $R_\perp$ have linear scaling with $\nu_s = \nu_\perp = 1$.
In this phase, because the walk is stretched perpendicular to the surfaces, the position of the endpoint in the plane parallel to the surface is not self-avoiding and so $\nu_\parallel = 1/2$.
This is the same behaviour as the endpoint-pulled model \cite{Bradly2023}.

In the compressed phase things are more complicated.
The bottom gray line marks the value $\nu_d/(\nu_d + 1)$, and for $y < 1$ the exponents $\nu_s$ and $\nu_\perp$ agree with this value as expected from the scaling arguments.
The agreement is better for $d = 2$ than $d = 3$ and there is also some systematic variance that is larger for $\nu_s$ than $\nu_\perp$, but overall we find that the scaling arguments match the numerical estimates for these two exponents.
The scaling of the endpoint height in the compressed phase is less clear.
For $d = 2$ the claim that $\nu_\parallel = 2\nu_d / (\nu_d + 1)$ (top gray line) is supported by the numerical estimates, but only when $y$ is sufficiently smaller than the critical point.
For $d = 3$ this does not appear to be the case; even allowing for some unknown systematic error it is not clear that $\nu_\parallel$ matches any expected value in the compressed region.
We report the value $\nu_\parallel \approx 0.6503(7)$ at $y = 0.1$ as indicative of the value in the compressed phase.

What is not shown in \fref{fig:Exponents} is the limit $y \to 0$ where the compression induces quasi $(d-1)$--dimensional behaviour.
At $y = 0$ the lattice is $d-1$ dimensional and we should see that $\nu_\parallel = \nu_{d-1}$ (e.g. $\nu_\parallel = 1$). 
Meanwhile, $\nu_s$ and $\nu_\perp$ are undefined at $y = 0$ since the span and endpoint height are identically zero.
However, we do not have sufficient accuracy to see if $\nu_\parallel$ discontinuously jumps to the lower-dimensional value or if there is a continuous crossover at very small $y$.

\subsection{Critical point scaling}
\label{sec:CriticalResults}

Finally, at the critical point $F = 0$ we estimate the critical exponents $\phi$ and $\alpha$.
Based on the scaling arguments from \sref{sec:Critical} we take the exponent estimate for $\phi = \nu_s$ from $s_n(0)$.
But we also estimate $\phi$ and $\alpha$ separately by fitting $c_n(F_\text{peak})$ with correction-to-scaling terms.
Our estimates for $\phi$ and $\alpha$ are listed in \ref{tab:CriticalExponents} for both $d = 2$ (square lattice) and $d = 3$ (simple cubic lattice), and are clearly consistent with $\phi = \nu_d$, and $\alpha < 1$ indicating a continuous transition.
The stated errors are only due to stochastic effects from the simulation and hence are quite small; systematic errors from higher-order corrections to leading order scaling are difficult to parse from Monte Carlo data beyond the next-to-leading order correction-to-scaling term in \eref{eq:SpanCorrectionToScaling}.

\begin{table}[t!]
\centering
	\begin{tabular}{l|c|c|c}
		$d$ 	& $\quad$ & $s_n(0) \sim S_0 n^{\phi-1}$ & $c_n(F_\text{peak}) \sim C_0 n^{2\phi-1}$ \\ \hline
		2		& $\phi$ 	& $ 0.7464(2) $ & $ 0.7520(1) $ \\ 
		(squ)	& $\alpha$ 	& $ 0.6603(4) $ & $ 0.6703(2) $ \\ \hline
		3		& $\phi$	& $ 0.58806(5) $ & $ 0.5900(4)  $ \\
		(sc)	& $\alpha$	& $ 0.2995(1) $ & $ 0.305(1) $
	\end{tabular}
	\caption{Critical exponents $\alpha$ and $\phi$ estimated from the scaling of thermodynamic quantities at the critical point $F_\text{c} = 0$ for pulled SAWs on the square and simple cubic lattices.} 
\label{tab:CriticalExponents}
\end{table}

\section{Conclusion}
\label{sec:Conclusion}

The application of a force to a lattice polymer via compression of two parallel plates has many similarities to the application of the force at the endpoint.
The critical point of zero force between compression and stretching displays the same critical phenomenon, namely the critical exponents are $\phi = \nu_d$ and thus $\alpha < 1$ indicating a continuous transition.

In the compressed phase we tested a prediction that $\nu_s = \nu_d / (\nu_d + 1)$, which is based on scaling arguments for compressed walks in two dimensions.
We find that this holds for $d = 2$, for both the scaling of the average span and the scaling of the end-to-end distance.
Although these arguments do not hold for $d = 3$, we find that this prediction for $\nu_s$ closely matches the estimated value from numerical simulations.
The numerical value is slightly higher than $\nu_3 / (\nu_3 + 1)$, which could be interpreted as being due to the walks retaining their 3D self-avoiding character in the compressed phase.
However, this is only speculative and we cannot rule out some systematic error or the effect of higher-order terms.
We note that equivalent simulations for compressed walks in 4D also support our findings.
For $d = 4$ the mean-field value of the metric exponent is $\nu_4 = 1/2$ and we find that our numerical estimates are consistent with $\nu_s = \nu_4 = 1/2$ at the critical point and $\nu_s = \nu_4 / (\nu_4 + 1) = 1/3$ in the compressed phase.
However, in addition to the same problems with the scaling arguments as 3D walks, there are also logarithmic corrections for $d = 4$ which complicates the scaling analysis and our data is not in a position to properly investigate.

\revision{Finally, we compare to what is known about the scaling of directed walk models with a compressing force.
Ref.~\cite{Beaton2015b} also considered scaling of simple random walks, which are relevant because the standard directed walk model for $d = 2$ (i.e.~Dyck paths) can be viewed as a one-dimensional simple random walk.
The prediction for this model is an exponent $\nu_s = 1/3$ in our notation.
This value is also predicted from the exact solution of a directed walk model, which although it is complicated for the compressed phase where $y < 1$, the asymptotic scaling can be extracted \cite{Guttmann2015}.
}

While the prediction from scaling arguments gave a good prediction for the leading order exponent $\nu_s$, it also predicts a higher order term $1/n$ for the average span in the compressed phase.
However, we found that this did not agree with our data, at least for the range of $n$ that we are able to simulate, and thus the predicted relation between the coefficient and the strength of the force does not hold.
Of course, it is not unusual to be unable to accurately detect higher-order scaling in Monte Carlo data, especially in addition to the slow convergence indicated by the presence of the exponential factor in \eref{eq:PartitionCompressedGeneral}.

\acknowledgements			
The authors acknowledge financial support from the Australian Research Council via its Discovery Projects scheme (DP230100674).
This research was supported by The University of Melbourne's Research Computing Services and the Petascale Campus Initiative.
Data generated for this study is available on request.


\IfFileExists{
	../../../../papers/bibtex/polymers_master.bib
}{
	\bibliography{../../../../papers/bibtex/polymers_master.bib}
}{
	\bibliography{compressed-saw.bib}
}

\end{document}